\documentclass[12pt]{article}

\usepackage{amsthm,amsmath,amssymb}
\usepackage{mathrsfs}
\usepackage{times}
\usepackage{epsfig,amsmath}
\usepackage{subfigure}
\usepackage{graphicx}
\usepackage{color}
\usepackage{epstopdf}
\usepackage{cite}
\usepackage{hyperref}
\usepackage{algorithm}
\usepackage{algorithmic}
\usepackage{mathrsfs}
\usepackage[version=4]{mhchem}
\usepackage{amsmath}
\usepackage{booktabs}
\usepackage{makecell}
\usepackage{array}
\usepackage{bm}
\usepackage{multirow}
\usepackage[resetlabels]{multibib}
\usepackage{hyperref}
\usepackage{url} 
\usepackage{caption}
\usepackage{siunitx}
\usepackage{comment}

\newenvironment{sciabstract}{%
\begin{quote} \bf}
{\end{quote}}

\newcounter{lastnote}

\title{A fully-programmable integrated photonic processor for both domain-specific and general-purpose computing}

\author
{Feng-Kai Han$^{1,2,\dagger}$, Xiao-Yun Xu$^{1,2,4,\dagger,\ast}$, Tian-Yu Zhang$^{1}$, Lei Feng$^{1,2}$,\\ Chu-Han Wang$^{1,2}$, Jie Ma$^{1,2}$, Ze-Feng Lan$^{1,2}$, Chao-Qian Li$^{1,2}$, Yi Xie$^{1,2}$,\\ Hai Yan$^{3}$, Yu-Fei Liu$^{3}$, Yu-Quan Peng$^{3}$, Xian-Min Jin$^{1,2,3,4,\ast}$\\
	\\
	\normalsize{$^1$Center for Integrated Quantum Information Technologies (IQIT), School of Physics}\\
	\normalsize{and Astronomy and State Key Laboratory of Photonics and Communications,}\\
	\normalsize{Shanghai Jiao Tong University, Shanghai 200240, China.}\\
	\normalsize{$^2$Hefei National Laboratory, Hefei, 230088, China}\\
	\normalsize{$^3$TuringQ Co., Ltd., Shanghai 200240, China}\\
	\normalsize{$^4$Chip Hub for Integrated Photonics Xplore (CHIPX), }\\
	\normalsize{Shanghai Jiao Tong University, Wuxi 214000, China}\\
	\normalsize{$^\dagger$These authors contributed equally.}\\
	\normalsize{$^\ast$E-mail: xiaoyunxu@sjtu.edu.cn}\\	
	\normalsize{$^\ast$E-mail: xianmin.jin@sjtu.edu.cn}\\	
}
\date{}

\begin{document} 
\baselineskip24pt

\maketitle

\begin{sciabstract}
A variety of complicated computational scenarios have made unprecedented demands on the computing power and energy efficiency of electronic computing systems, including solving intractable nondeterministic polynomial-time (NP)-complete problems and dealing with large-scale artificial intelligence models. Optical computing emerges as a promising paradigm to meet these challenges, whereas current optical computing architectures have limited versatility. Their applications are usually either constrained to a specialized domain or restricted to general-purpose matrix computation. Here, we implement a fully-programmable integrated photonic processor that can be configured to tackle both specific computational problems and general-purpose matrix computation. We achieve complete end-to-end control of the photonic processor by utilizing a self-developed integrated programmable optoelectronic computing platform. For domain-specific computing, our photonic processor can efficiently solve two kinds of NP-complete problems: subset sum problem (far more than $2^N$ different instances) and exact cover problem. For general-purpose computation, we experimentally demonstrate high-precision optical dot product and further realize accurate image edge detection and MNIST handwritten image classification task with an accuracy of 97\%. Our work enhances the versatility and capability of optical computing architecture, paving the way for its practical application in future high-performance and complex computing scenarios.

\end{sciabstract}

\subsection*{Introduction}
Conventional electronic computers with the von Neumann architecture are faced with significant practical limitations, failing to meet the growing requirement for computational capabilities. On the one hand, physical limitations such as quantum tunneling effect fundamentally hinder the further progression of Moore’s law\cite{Waldrop2016chips}. On the other hand, the von Neumann architecture inherently suffers from the separation of memory and processing units, leading to severe data transfer latency and extra energy consumption\cite{Theis2017theend}. Moreover, various applications, ranging from training trillion-parameter artificial intelligence models to solving intractable NP-complete problems, demand computing architecture innovation. These challenges have driven extensive research into novel computing architectures including optical computing\cite{Lin2024120,Ouyang202516}, biological computing\cite{Adleman1994Molecular,Braich2002Solution}, and quantum computing\cite{Farhi2001quantum,Gao2022quantum}.

Optical computing possesses high speed, massive parallelism, and low energy consumption due to the intrinsic properties of photons\cite{xu2023Integrated}. These advantages make optical computing a promising paradigm to fulfill the growing needs of computational power and energy efficiency\cite{Lin2018all,Zhou2021Large}. Traditional optical computing system, which is typically implemented with bulk optical components, is limited in miniaturization and scalability. In recent years, integrated photonics have made chip-scale and scalable photonic processor possible\cite{Venema2011sillicon}. With the advancement of integrated photonic technologies, photonic processors have ushered in a new era of information processing\cite{Shekhar2024Roadmapping,Seok2019Wafer}, emerging as a major building block in high-performance computing. Diverse optical computing architectures have been adopted in integrated photonic processors to accelerate matrix computation, including reconfigurable Mach-Zehnder interferometer (MZI) meshes \cite{shen2022deep,zhang2021,xu2021optical,zhang2022low,li2024High-efficiency}, wavelength-division multiplexed microring banks \cite{Tait2017Neuromorphic, Feldmann2019all,huang2021sillicon}, on-chip diffractive metasurfaces\cite{Zarei2020Integrated,fu2023photonic}, and other integrated optical components \cite{Ashtiani2022an,dong2023higher,zhu2022space,Zelaya2025Integrated}. 

In addition, optical computing enables to rapidly solve NP-complete problems which are computationally hard for electronic computers \cite{Sipser2012Introduction}. According to computational complexity theory \cite{Garey1979computers}, the solution space of NP-complete problems grows super-polynomially with the problem size. Meanwhile, NP-complete problems have widespread applications in transportation, industrial manufacturing, finance, etc. Therefore, it is of great practical significance to solve NP-complete problems efficiently \cite{Biesner2022solving,Eghdami2021SSA,Smith2005Fault}. It should be stressed that the computation of NP-complete problems differs from matrix computation. The computing architectures for NP-complete problems are usually specially-designed and exhibit features distinct from matrix computation \cite{Farhi2001quantum,chang2018quantum,Adleman1994Molecular,Oltean2009solving,Oltean2008solving,Vazquez2018optical,tan2023scalable,Gao2024ising,Haist2007an,wu2014optical,xu2020scalable,jiang2023programmable,xu2024Reconfigurable}. We notice that existing optical computing architectures generally focus on a single category of computational problems and have very limited application range. There is no such unified integrated optical computing platform capable of dealing with both NP-complete problems and matrix-based computation (e.g., convolutions in neural networks), which indicates the lack of versatility of current optical computing architectures.

In this work, we present a fully-programmable integrated photonic processor adaptable to both domain-specific (i.e., NP-complete problems) and general-purpose computing (i.e., matrix multiplications). We demonstrate the capability of the photonic processor to solve two different kinds of NP-complete problems, subset sum problem and exact cover problem. By precisely reconfiguring our photonic processor, more than $2^N$ different subset sum problem instances and various exact cover problem instances have been efficiently solved with 100\% accuracy. Furthermore, without modifying the underlying photonic hardware, we are able to perform neural network tasks with high precision, including image edge detection and MNIST handwritten image classification. Our investigation confirms the feasibility of tackling two distinct categories of computational tasks using the same integrated photonic architecture, further improving the universal applicability of integrated optical computing.

\subsection*{Results}
\begin{figure}[htbp]
	\centering
	\includegraphics[width=1.0\linewidth]{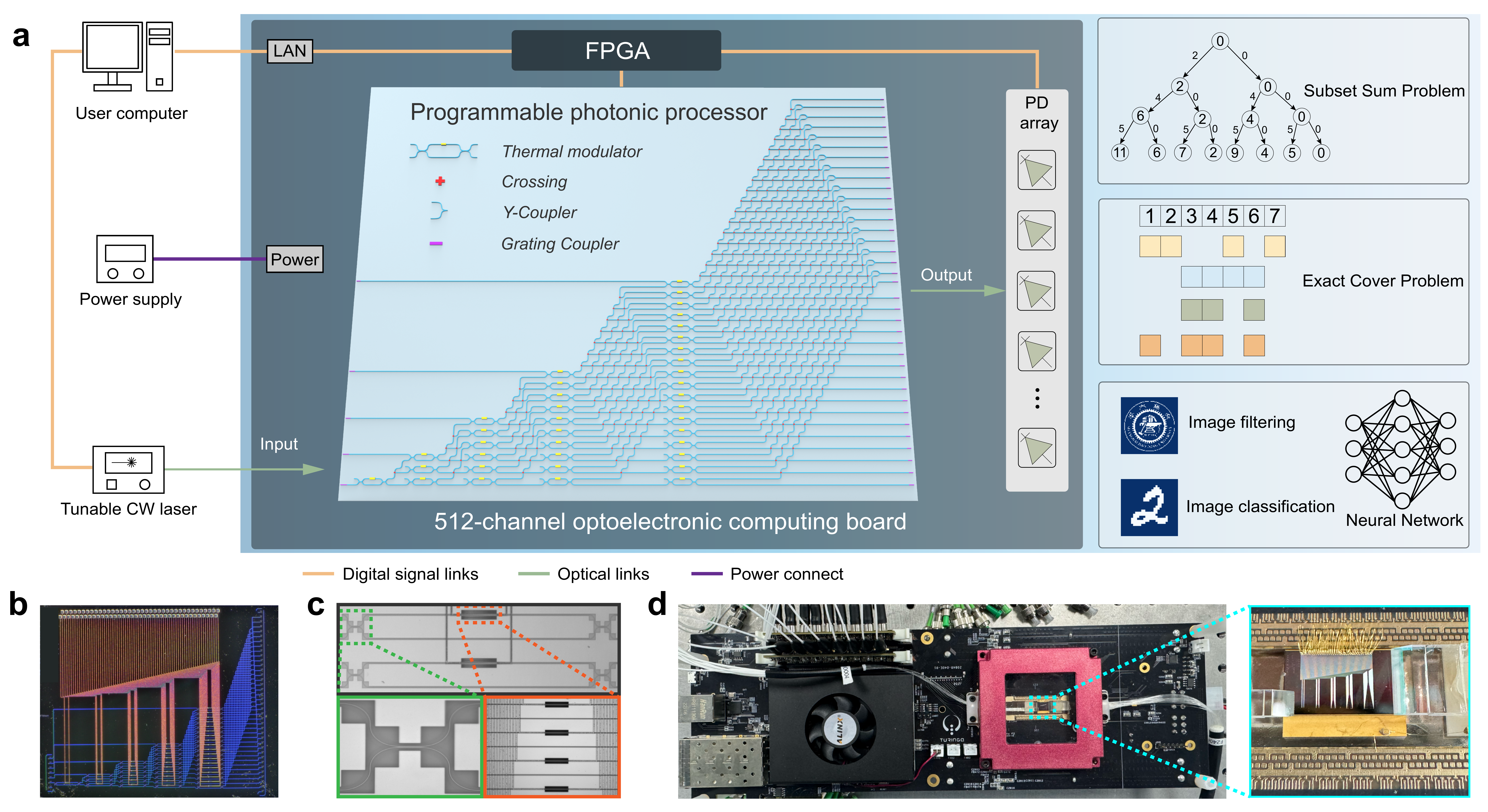}
	\caption{\textbf{The fully programmable integrated photonic processor and experimental setup.} \textbf{a} {The integrated photonic processor is composed of four kinds of standardized optical components, and it is integrated onto a 512-channel optoelectronic computing board which incorporates an FPGA, a TEC and a PD array. Users can use an external computer to program the photonic processor and acquire the collected data through a LAN  port connected to the optoelectronic computing board. The whole optoelectronic computing system is driven by a tunable CW laser and a programmable power supply. The photonic processor can be applied to different computing scenarios: solving the subset sum problem and exact cover problem, and conducting matrix multiplications and convolutions in neural networks.} \textbf{b} {Optical microscopic image of the fabricated photonic chip.} \textbf{c} {Images of the on-chip MZIs, directional coupler (encircled by green lines), and electrodes with air trenches on both sides (encircled by orange lines).} \textbf{d} {Photograph of the optoelectronic computing system. Inset on the right shows the packaging of the photonic chip.}}

	\label{f1}
\end{figure}

As illustrated in Fig. \ref{f1}a, the integrated photonic processor is composed of four types of standardized optical components (a total of 498), which are arranged in a particular manner to form a functional network (the microscopy image of the photonic chip is shown in Fig. \ref{f1}b). Light is coupled into and out of the photonic processor through grating couplers. Thermally modulated MZI acts as the basic modulation unit, which uses directional couplers as splitters to reduce its footprint as marked by green lines in Fig. \ref{f1}c, thereby enhancing the scalability of the photonic processor. Additionally, we implement air trenches on both sides of the electrodes in the MZIs as denoted by orange lines in Fig. \ref{f1}c, which is beneficial to reducing thermal crosstalk and improving modulation efficiency. The tunable MZIs together with the five optional input channels endow the photonic processor with programmability.

We integrate the photonic chip onto a home-made 512-channel optoelectronic computing board to realize precise and stable end-to-end control (the photograph is shown in Fig. \ref{f1}d), which incorporates a multi-channel field programmable gate array (FPGA), temperature controller (TEC) and a photodetector (PD) array (see Methods for details). The computing board is connected to an external computer through a local area network (LAN) port, which provides users an interface to program the photonic processor and acquire the collected data. The whole optoelectronic computing system is driven by a tunable continuous-wave (CW) laser and a programmable power supply (see Methods for details). Leveraging the full programmability and the ingenious architecture design, our integrated photonic processor is able to perform various computational tasks belonging to different categories: (a) solving the subset sum problem and (b) exact cover problem, and (c) conducting matrix multiplications and convolutions in neural networks, as depicted in the rightside of Fig. \ref{f1}a.

\subsubsection*{Subset sum problem solver}
We experimentally configure the photonic processor for domain-specific problems. Different kinds of NP-complete problems are successfully mapped onto the photonic processor, without modifying the hardware architecture. Subset sum problem is one of the NP-complete problems to solved. Given a set $S$ containing $N$ integers, the subset sum problem asks whether there exists a subset of $S$ whose sum equals a target $T$. It is a decision problem, meaning that the answer to the problem is either yes or no. 

\begin{figure}[htbp]
	\centering
	\includegraphics[width=1.0\linewidth]{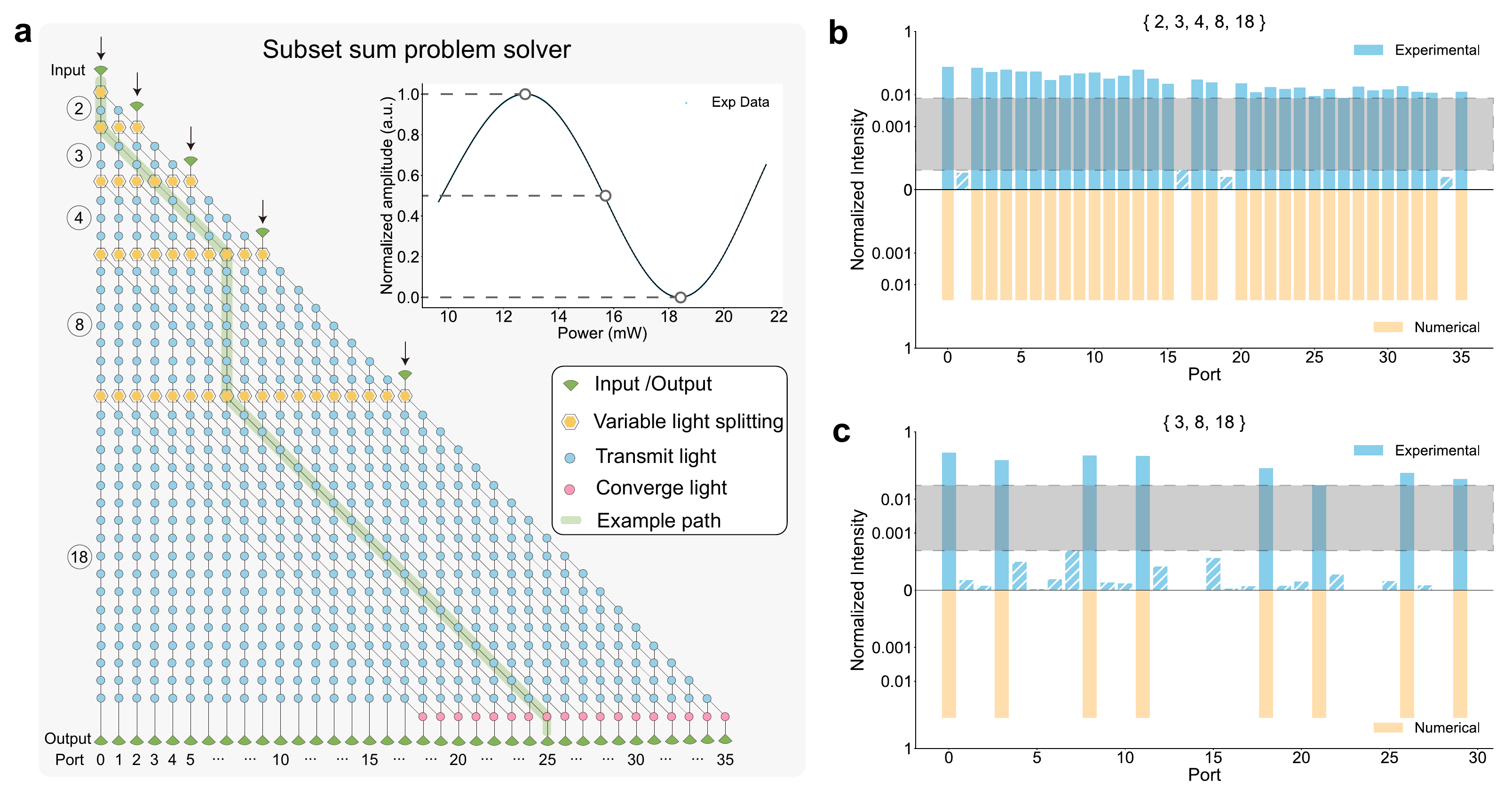}
	\caption{\textbf{Architecture and computing results of the subset sum problem solver.} \textbf{a} {The photonic architecture adapting to solving the subset sum problem can be represented by the abstract network made of lines (optical path) and four kinds of nodes (optical component). The integers on the left side correspond to the elements in the set $S$, which equals the vertical distance between two adjacent rows of hexagonal nodes. The green path represents the selection of elements 3, 4, and 18, and the sum is 25 as denoted by the output port number. The inset shows one of the typical characterization curves of the on-chip MZIs (yellow hexagonal nodes).} \textbf{b} {Computing results of the case $\{2, 3, 4, 8, 18\}$. The tolerance interval for the threshold, which classifies the outputs into valid signals and invalid noise (highlighted with white diagonal stripes), has an upper bound of 0.00787 and a lower bound of 0.00003.} \textbf{c} {Computing results of the case $\{3, 8, 18\}$. The threshold has an upper bound of 0.02609 and a lower bound of 0.00029.}}

	\label{f2}
\end{figure}

The photonic architecture adapting to solving the subset sum problem can be represented by an abstract network composed of lines and nodes, as shown in Fig. \ref{f2}a. The lines denote optical paths. The four kinds of nodes represent the  input/output, and the modules that function as splitting light, transmitting light and converging light, respectively. Light is injected through one of the green sector nodes (grating couplers in Fig. \ref{f1}a). At yellow hexagonal nodes (thermally modulated MZIs in Fig. \ref{f1}a), light can be split with any specified ratio $\eta:1-\eta$ ($0 \leq \eta \leq 1$) by appropriately setting the applied current. Then the split light propagates vertically and diagonally. All the MZIs have been characterized, and the experimental results excellently agree with theory, validating the precise phase control capability. One of the typical characterization curves is displayed in the inset of Fig. \ref{f2}a. Based on the curves, we can identify three critical working states of the MZIs (as denoted by hollow circles): bar state ($\eta=1 $), cross state ($\eta=0$), and balanced state ($\eta=0.5 $). Blue circular nodes (crossings or waveguide segments in Fig. \ref{f1}a) enable light to move forward along original direction. At the end of the network, pink circular nodes (Y-couplers in Fig. \ref{f1}a) gather light from different paths.

We map the subset sum problem to the photonic processor according to the following rules\cite{xu2024Reconfigurable}. First, a yellow hexagonal-node block and a blue circular-node block are alternately arranged for $N$ iterations (here, $N=5$). Second, the vertical distance between two adjacent rows of hexagonal nodes corresponds to the elements in the set $S$, as indicated by the integers on the left side in Fig. \ref{f2}a. The distance is quantified by the number of nodes. Third, diagonal propagation of light indicates the inclusion of an element into the summation, whereas vertical propagation indicates the exclusion. Finally, the position of the output signals represents the resultant sums, which are identified by the output port numbers. For instance, the path highlighted in light green demonstrates the inclusion of elements 3, 4, and 18, and the exclusion of elements 2 and 8, yielding a summation of 25. In the photonic processor, the incident laser explores all the possible optical path, inherently searching for the solution at light speed. The presence or absence of optical signal at the target output port gives a yes or no answer. The photonic architecture allows single-shot measurement across all output ports. 

There are two approaches to program the photonic processor, i.e., switching the input channel or changing the working states of the MZIs. Our programming scheme allows any combination of the $N$ elements in the set $S$, which enables the mapping of $2^N$ different subset sum problem instances (see Methods for details). Besides, an addition of adjacent elements in the set $S$ could generate a new element. Under this condition, more subset sum problem instances containing the new elements can be mapped to the photonic processor (see Methods for details). For our fully reconfigurable photonic processor, it supports 89 (i.e., $2^5+57$) different configurations for the subset sum problem.

The photonic processor is first configured to solve the subset sum problem instance where the set $S=\{2,3,4,8,18\}$. In the theoretical framework, any output signal with nonzero intensity indicates the existence of the corresponding subset sum. However, in the experiments, this is not necessarily the case due to unavoidable environmental noise and fabrication imperfections. We employ a reasonable intensity threshold to accurately classify the experimental signals into valid and invalid certifications. As shown in Fig. \ref{f2}b, the signals with intensities exceeding the threshold are identified as valid, whereas those below the threshold are invalid (highlighted with white diagonal stripes). Our photonic processor achieves 100\% accuracy compared to the theoretical predictions. As indicated by the gray shaded region, the tolerance interval for the threshold has an upper bound of 0.00787 and a lower bound of 0.00003, showing a good signal-to-noise ratio. By tuning the working states of particular MZIs, we achieve to address other subset sum problem instances. As shown in Fig. \ref{f2}c and Supplementary Section 1, in all the cases, the experimental results are in accordance with the theoretical results, demonstrating the reliable reconfigurability and high accuracy of our photonic processor.

\subsubsection*{Exact cover problem solver}
Except for solving the subset sum problem, our photonic processor can be further applied to another NP-complete problem, exact cover problem. The exact cover problem is also a decision problem \cite{Oltean2008exact} , with applications in airline fleet optimization and cloud resource allocation. It is formally defined as: given a collection $S$ of subsets, each containing elements of a target set $X$, exact cover problem asks whether a subcollection $S^*\subseteq S$ exists such that each element in $X$ is contained in exactly one subset in $S^*$. For example, let $S_1=\{[1,2],[2,3],[3],[3,4]\}$ be a collection of subsets and a target set $X_1=\{1,2,3,4\}$, the subcollection $\{[1,2],[3,4]\}$ is an exact cover of $X_1$ while $\{[1,2], [3], [3,4]\}$ is not an exact cover. 

\begin{figure}[htbp]
	\centering
	\includegraphics[width=1.0\linewidth]{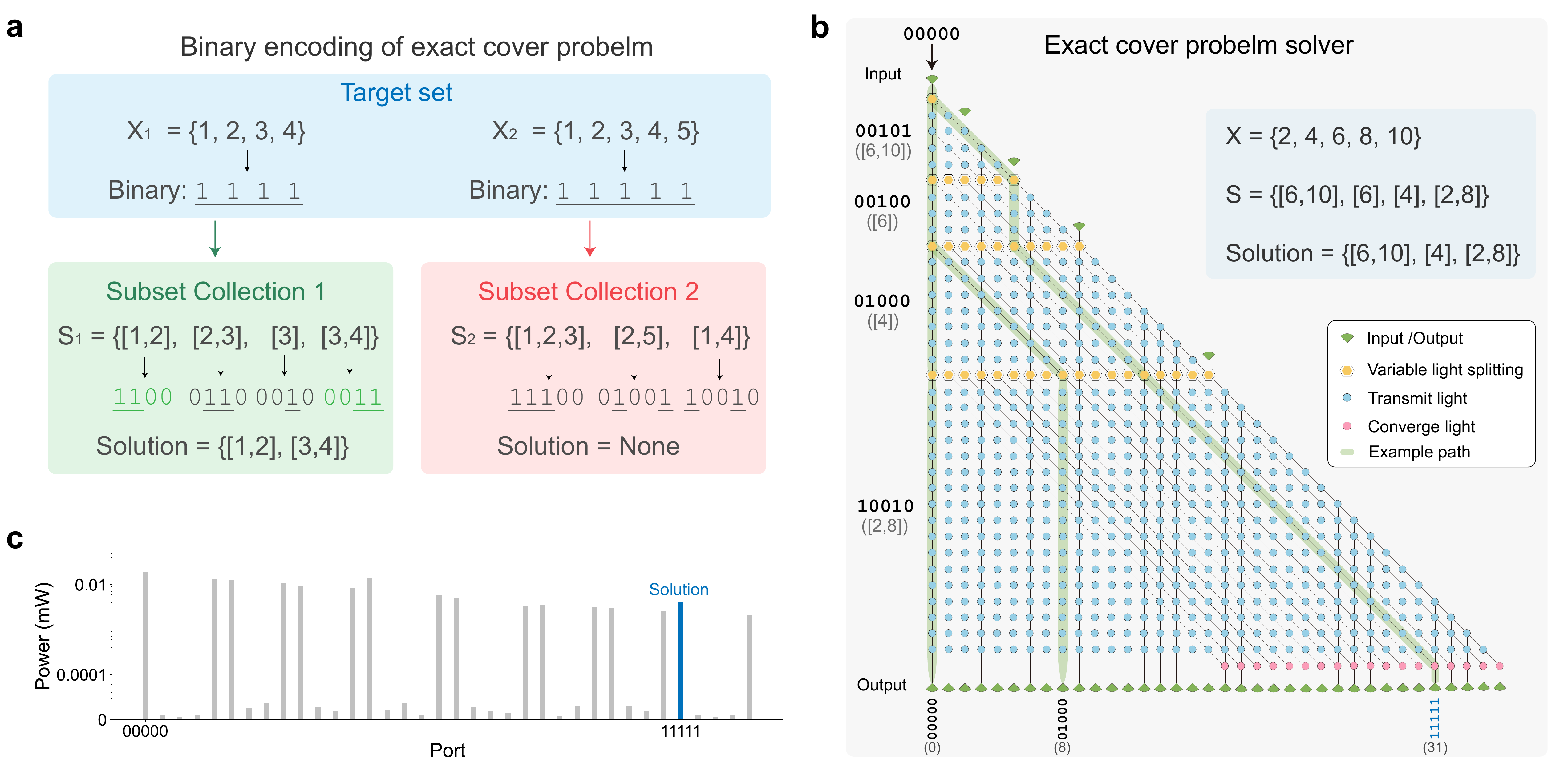}
	\caption{\textbf{Architecture and computing result of the exact cover problem solver.} \textbf{a} {Binary encoding scheme for the exact cover problem. For instance, target set $X_1$ is encoded as $1111$, subset collection $S_1$ is encoded as $\{1100, 0110, 0010, 0011\}$. For a 4-element target set, an exact cover exists if there is a subcollection whose subsets have the bitwise sum $1111$. The subcollection $\{[1,2],[3,4]\}$ is an exact cover of $X_1$. No exact cover of $X_2$ exists in $S_2$.} \textbf{b} {The exact cover problem solver has the same photonic architecture as the subset sum problem solver while some of the mapping rules are different. The binary number on the leftside, corresponding to the subsets in $S$, equals the vertical distance between two adjacent rows of hexagonal nodes. A detection of light at output port identified by the binary number $11111$ indicates that an exact cover exists.} \textbf{c} {Experimental result for the exact cover problem instance where target set $X=\{2, 4, 6, 8, 10\}$ and subset collection $S=\{[6,10],[6],[4],[2,8]\}$. The detection of output signal at output port $11111$ (blue bar) represents the existence of an exact cover.}}
	
	\label{f3}
\end{figure}

To map the exact cover problem onto our photonic processor, we first employ a binary scheme to encode the exact cover problem\cite{Korten2021design}. As illustrated in Fig. \ref{f3}a, target set $X_1=\{1,2,3,4\}$ is encoded as $1111$, subset collection $S_1=\{[1,2],[2,3],[3],[3,4]\}$ is translated as binary code $\{1100, 0110, 0010, 0011\}$. When a target set contains 4 elements, an exact cover exists if there is a subcollection whose subsets have the bitwise sum $1111$. Obviously, the subcollection $\{[1,2](1100),[3,4](0011)\}$ is an exact cover of $X_1$. In contrast, for target set $X_2=\{1,2,3,4,5\}$ and subset collection $S_2=\{[1,2,3],[2,5],[1,4]\}$ (encoded as $\{11100, 01001, 10010\}$), none of the subcollections has the bitwise sum $11111$, indicating that there is no exact cover. The photonic architecture adapting to binary-encoded exact cover problem is the same as the subset sum problem, as exhibited in Fig. \ref{f3}b. The definition of including (excluding) a subset into (out of) the summation is similar to the subset sum problem solver. Whereas, other mapping rules have changed. (1) The vertical distance between two adjacent rows of yellow hexagonal nodes corresponds to the binary codes of the subsets in the collection $S$, as indicated by the binary number on the leftside of Fig. \ref{f3}b. (2) A detection of light at output port identified by the binary number $11..1$ (the number of $1$ equals the number of elements in target set $X$) indicates that an exact cover exists.

Using the binary encoding scheme and the updated mapping rules, we experimentally solve the exact cover problem instance where target set $X=\{2,4,6,8,10\}$ and subset collection $S=\{[6,10],[6],[4],[2,8]\}$ (encoded as $\{00101, 00100, 01000, 10010\}$), as illustrated in Fig. \ref{f3}b. The detection of output signal at port $11111$ (blue bar in Fig. \ref{f3}c) represents the existence of an exact cover, confirming our photonic processor’s capability to precisely tackle exact cover problem at light speed. The photonic processor can further solve other exact cover problem instances which have a different-sized target set or a different subset collection (see Methods for details). In all the cases, the experimental results show 100\% accuracy, which are provided in Supplementary Section 2. The successful application on the two kinds of NP-complete problems validates the versatility and adaptability of the optical computing architecture.

\subsubsection*{Photonic dot-product engine}
\begin{figure}[htbp]
	\centering
	\includegraphics[width=1.0\linewidth]{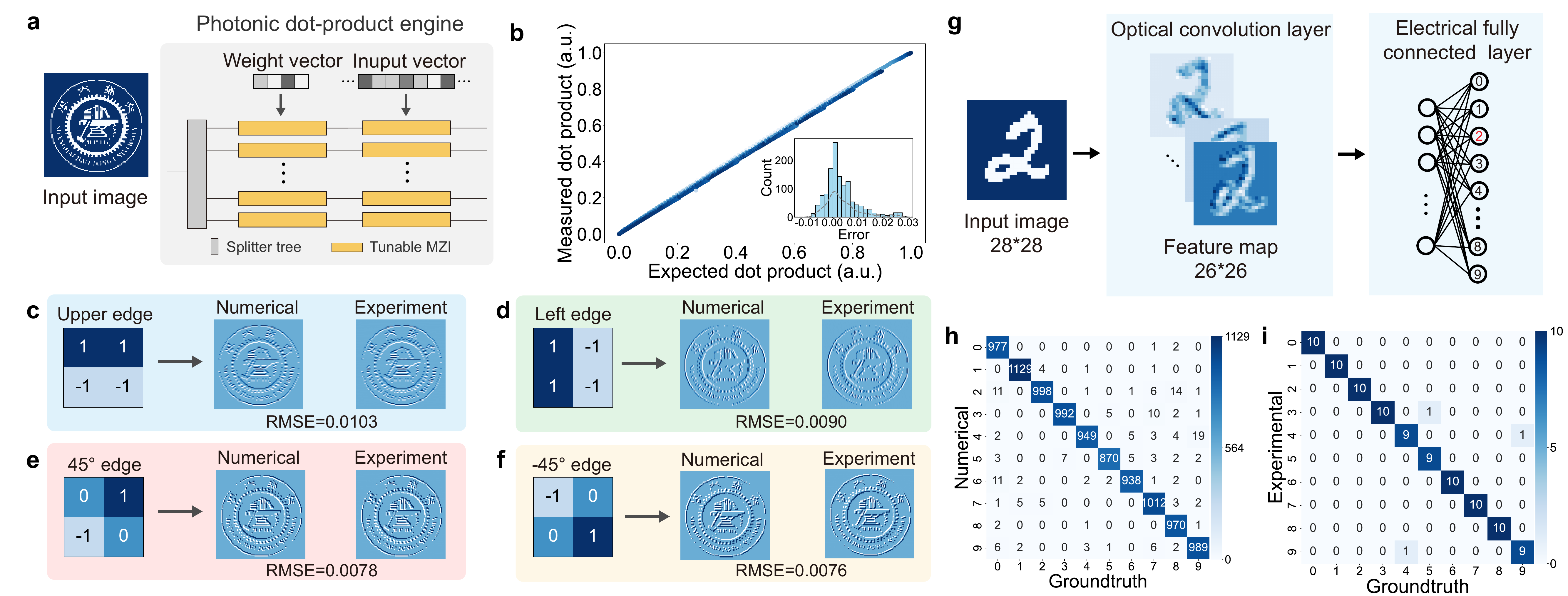}
	\caption{\textbf{Architecture and results for the photonic dot-product engine.} \textbf{a} {The architecture of photonic processor adapted to conducting optical dot product is composed of a splitter tree and cascaded MZIs. The input vector of the image and weight vector are encoded onto two MZIs in the same row. The matrix multiplication is implemented by row-wise encoding of the weight matrix onto the corresponding column of MZIs.} \textbf{b} {Experimental results of 1300 random multiplication. The experimental results are in good agreement with the theoretical results and the experimental error has a small standard deviation of 0.0067 as displayed in the inset.} \textbf{c, d, e, f} {The numerical and experimental results of image edge detection with four 2 × 2 kernels (upper edge, left edge, 45$^\circ$ edge and -45$^\circ$ edge).} \textbf{g} {The main structure of the convolutional neural network for image classification, which contains an optical convolution layer and electrically fully connected layers.} \textbf{h, i} {Confusion matrices of MNIST dataset classification for the numerical model and our photonic processor.}}
	\label{f4}
\end{figure}

Beyond addressing domain-specific computational problems, our photonic processor is also adaptable to general-purpose matrix computing. Since the dot product is a building block of matrix multiplications or convolutions, which are the core of artificial neural networks (ANNs)\cite{Rawat2017deep,Abdel2014Convolutional,Sladojevi2016deep,Miglani2019deep,Fu2025Passive}, constructing a photonic dot-product engine offers a way to accelerate matrix-vector multiplications of ANNs by harnessing the advantages of light-speed parallel processing. As introduced above, we can precisely control the propagation path of light in the photonic processor. Through ingenious setting of the working states of the MZIs, our photonic processor can be reconfigured to a typical cascaded MZI network which enables optical dot product\cite{li2024High-efficiency,Xu2021Optical} (see Methods and Supplementary Section 3 for details), as depicted in Fig. \ref{f4}a. In optical dot product, two operands are encoded onto the phase shifters of two cascaded MZIs respectively, and the output optical power represents their product. The cascaded MZI configuration can further perform dot product of weight vector \boldsymbol{$w$} and input vector \boldsymbol{$x$} via temporal multiplexing. Namely, the elements in weight vector \boldsymbol{$w$} (input vector \boldsymbol{$x$}) are sequentially encoded onto the first (the second) MZI.

The above principle can be extended to matrix multiplication by row-wise encoding of the weight matrix. Specifically, the dot product operation \boldsymbol{$y=w\cdot x$} can be generalized to matrix multiplication $Y=W\cdot X$ by encoding each row of the weight matrix $W$ onto the corresponding MZIs in the first column in Fig. \ref{f4}a (e.g., the $i$-th row of $W$ is encoded onto the $i$-th MZI in the first column). On this basis, our implemented photonic processor facilitates up to 8-dimensional optical dot product calculations. We experimentally test 1300 random multiplications (scaled between 0 and 1), which are in good agreement with the theoretical results as shown in Fig. \ref{f4}b. The experimental error has a small standard deviation of 0.0067 which is equivalent to a computing accuracy of 7.22 bits, as displayed in the inset of Fig. \ref{f4}b, confirming the photonic processor’s high precision in performing general-purpose matrix computation.

We have further implemented image edge detection of a 200 × 200 grayscale pixel matrix of the Shanghai Jiao Tong University emblem (the input image in Fig. \ref{f4}a) with the photonic processor. Four parallel 2 × 2 convolution operations are carried out for a comprehensive edge detection of the image. The experimental results (Figs. \ref{f4}c-f) exhibit sharp feature extraction and possess a low average root mean square error (RMSE) of 0.0087 compared to numerical simulation. The achievement of simultaneously performing multi-kernel convolution validates the capability of our photonic processor in multi-dimensional dot product calculations. To further showcase the processor’s versatility, we construct a hybrid photonic-electronic convolutional neural network for MNIST handwritten digit classification. As illustrated in Fig. \ref{f4}g, the photonic processor is used for the optical convolution layer while the electronic computer is responsible for the fully connected layers. We have achieved a classification accuracy of 97\%, which closely agree with the numerical result of 97.5\%, as shown by the confusion matrices in Fig. \ref{f4}h and Fig. \ref{f4}i. The performance of our photonic processor in executing the two aforementioned tasks testifies its high reliability in real-world machine learning applications, which further confirm that the adaptability of our optical computing architecture is beyond NP-complete problems.

\subsection*{Discussion}
In summary, we present a fully-programmable integrated photonic processor supporting both domain-specific and general-purpose matrix computation. The photonic processor is fabricated via a CMOS-compatible process and integrated with a custom-designed multi-channel optoelectronic computing board to form a fully programmable optoelectronic computing system. Based on this architecture, we have carried out four tasks which belong to two distinct categories: NP-complete problems (subset sum problem and exact cover problem) and matrix-based computation tasks (image edge detection and MNIST handwritten image classification). In the case of NP-complete problems, our photonic processor has efficiently and accurately solved more than $2^5$ different subset sum problem instances and various exact cover problem instances. Our approach leverages light-speed parallel exploration to enable ultrafast optical computation, showing a potential speedup in computing time (see Supplementary Section 5).  In the case of matrix-based computation tasks, our photonic system facilitates energy-efficient matrix computation with ultra-low latency. On this basis, we further realize accurate image edge detection and MNIST handwritten image classification task with 97\% accuracy. Photonic techniques such as wavelength-division multiplexing (WDM) or heterogeneous integration could be employed in the future to take full advantage of our optical computing architecture\cite{Ou2024Hypermultiplexed,Churaev2023heterogeneously}.

The integrated photonic processor is experimentally demonstrated for the first time to deal with both NP-complete problems and general-purpose matrix computation. Thanks to the full programmability, the versatility of the photonic processor has been greatly enhanced, which could play a key role in future supercomputing. First, many real-life problems and algorithms based on NP-complete problems\cite{Schwerdfeger2016fast,Alonistiotis2024Approximating}, which typically demand programmable hardware, can be effectively addressed within the framework of our optical computing architecture. Second, our integrated photonic processor has the potential of solving a broader range of real-world neural network applications such as natural language processing, autonomous systems, and real-time translation\cite{Bahdanau2014Neural,Voulodimos2018deep}. Last but not least, our optical computing architecture provides a possible platform for hybrid computational scenarios. For instance, financial auditing task contains handwritten invoice recognition (formulated as an image classification task based on neural network) and consistency check of financial documents (formulated as subset sum problem-based computation)\cite{Biesner2022solving}. Given that the two subtasks should be executed collaboratively, conventional optical computing architecture would require separate domain-specific and general-purpose solvers, which results in an increase of system complexity and power consumption. In contrast, our photonic processor could provide a unified platform to meet the challenges through dynamic reconfiguration, showing exceptional adaptability for real-world complicated computing scenarios.

\subsection*{Methods}

\paragraph*{Fabrication and packaging of the integrated photonic processor.} Our photonic chip is fabricated using standard CMOS processes on a silicon nanophotonics platform. The silicon waveguide has dimensions of 220 nm × 450 nm to ensure single mode. The photonic chip features a compact footprint of 6 mm × 5 mm. The propagation loss of the waveguide is approximately 1.5 dB/cm. The insertion loss of the crossing is less than 0.2 dB. To enable efficient optical coupling, grating couplers in the integrated photonic processor are packaged with a 7-channel standard polarization-maintaining fiber array for input and a 38-channel single-mode fiber array for output, with a coupling loss of 4 dB per facet. The 72 electrical pads in the photonic chip are wire-bonded to a printed circuit board (PCB) and connected to a home-made 512-channel optoelectronic computing board (with a maximum refresh rate of 100 kHz) for controlling the power of the thermal phase shifters of the 38 on-chip MZIs. The MZIs are tuned using 100-$\mu$m-long thermal phase shifters, which modulate the refractive index of the waveguide through localized heating. The photonic chip is stably maintained at room temperature by a dedicated TEC (see Supplementary Section 4 for details).

\paragraph*{Experimental setup.}The light source is a C-band tunable CW laser with a maximum power of 12 dBm (Santec TSL-710). The 1552 nm laser transmits through a fiber polarization controller and then is coupled into the integrated photonic processor through fiber array. The output signals are detected by a pigtail positive-intrinsic-negative (PIN) photodetector array and collected by a data acquisition module on the optoelectronic computing board. For accurate characterization, a high-precision multi-channel optical power meter (Santec MPM-210) is employed to calibrate the MZIs. All measurements are conducted under standard room ambient conditions.

\paragraph*{Programming the subset sum problem solver.} Different subset sum problem instances can be solved by reconfiguring the photonic processor. We can realize 89 (i.e., $2^5+57$) different configurations of the subset sum problem solver, each of which corresponds to an individual subset sum problem instance. In a general case, the solver is initially designed for an subset sum problem instance where $S = \{ x_1,\ x_2,\ \dots,\ x_N \}$. There are two approaches to programming the solver. First, by switching to a different input channel, such as input channel $i$, the subset sum problem solver can be programmed to solve the subset sum problem instance where $S = \{ x_i,\ x_{i+1},\ \dots,\ x_N \}$. This is achieved by bypassing the first $i-1$ elements in the local network, effectively excluding these elements from the computation. Second, the inclusion or exclusion of a specific element $x_j$ can be controlled by appropriately setting the working states of the $j$th row of MZIs. By configuring the $j$th row of MZIs to bar state ($\eta=1 $) or cross state ($\eta=0 $), depending on their specific location, the light can be entirely directed to vertical paths. This ensures a zero probability of including the element $x_j$ in any summation, effectively removing it from the computation. Conversely, $x_j$ is retained when the MZIs are in balanced state ($\eta=0.5 $). In summary, the tunnable MZIs provide the capability to decide whether to include or exclude an element, thereby allowing all the possible combinations of the $N$ elements in the set $S$ (i.e., enabling the mapping of $2^N$ different subset sum problem instances).

Furthermore, the addition of adjacent elements within the set $S$ can generate an element which does not belong to $S$. For instance, we can achieve an addition of element $x_{j-1}$ and $x_j$ by setting the leftmost MZI in the $j$-th row to bar state (light propagates vertically) and the other MZIs in the $j$-th row to cross state (light propagates diagonally). Thereby, a new set $S' = \{x_1,\ \dots,x_{j-1}+x_j,\ \dots,\ x_N \}$ can be mapped to the subset sum problem solver. This scheme can add 57 different configurations, which further extends the programming ability of the subset sum problem solver.

\paragraph*{Programming the exact cover problem solver.}Different exact cover problem instances can also be mapped by reconfiguring the photonic processor. As the abstract network made of lines and nodes in Fig. \ref{f3}b shows, a target set $X$ with up to five elements can be mapped to the exact cover problem solver. The size of target set of exact cover problem instances mapped to the solver can be modified. To realize a target set with four elements, MZIs in the fifth row are set to cross state. To realize a target set with three elements, MZIs in the fourth and the fifth rows are set to cross state.

Besides, a change of subset collection also corresponds to a different exact cover problem instance. By summing two adjacent binary number on the leftside of Fig. \ref{f3}b, we can generate a new binary number, which provides a possibility to mapping a different subset collection. For instance, $00010$ and $00011$ can be summed to generate a new binary number $00101$ by setting the leftmost MZI in the second row to bar state and the other MZIs in the second row to cross state. Generally speaking, a new subset collection could be mapped to the exact cover problem solver using the addition of binary number $x_{j-1}$ and $x_j$, which is realized by setting the leftmost MZI in the $j$-th row to bar state (light propagates vertically) and the other MZIs in the $j$-th row to cross state (light propagates diagonally).

\paragraph*{Configuring the photonic processor to dot-product engine.} Through setting the working states of MZIs, we can configure our photonic processor to the architecture composed of a splitter tree and cascaded MZIs (see Fig. S7 in Supplementary Section 3), which can perform multi-dimensional optical dot product. For an 8-dimensional dot product operation, light is injected through the first input channel. By setting the first three rows of MZIs to balanced state while allowing the last two rows of MZIs to be freely programmed, we implement a splitter tree at the front end. Each branch of the splitter tree is connected to two cascaded MZIs, which are used for the encoding of two operands in optical dot product. For a 4-dimensional dot product operation, light is injected through the second input channel, and the MZIs in the second and third rows are set to balanced state and the remaining MZIs are left freely programmed.

\subsection*{Data availability}
The data that supports the findings of this study are available from the corresponding authors upon reasonable request.

\subsection*{Code availability}
The code used in this study is available from the corresponding authors upon reasonable request.

\clearpage

\subsection*{Acknowledgments}
This research is supported by the National Key R\&D Program of China (Grants No. 2024YFA1409300, No. 2019YFA0308703, No. 2019YFA0706302, and No. 2017YFA0303700); National Natural Science Foundation of China (NSFC) (Grants No. 62235012, No. 11904299, No. 61734005, No. 11761141014, and No. 11690033, No. 12104299, and No. 12304342); Innovation Program for Quantum Science and Technology (Grants No. 2021ZD0301500, and No. 2021ZD0300700); Science and Technology Commission of Shanghai Municipality (STCSM) (Grants No. 20JC1416300, No. 2019SHZDZX01, No.21ZR1432800, No. 22QA1404600, No. 24ZR1438700, No. 24ZR1430700 and No. 24LZ1401500); Shanghai Municipal Education Commission (SMEC) (Grants No. 2017-01-07-00-02-E00049); China Postdoctoral Science Foundation (Grants No. 2020M671091, No. 2021M692094, No. 2022T150415); Startup Fund for Young Faculty at SJTU (SFYF at SJTU)(Grants No. 24X010502876 and No. 24X010500170). X.-M.J. acknowledges additional support from a Shanghai talent program and support from Zhiyuan Innovative Research Center of Shanghai Jiao Tong University.

\subsection*{Author contributions}
{X.-Y.X. and X.-M.J. conceived the idea and supervised the project. F.-K.H., and X.-Y.X. implemented the algorithm and conducted the numerical experiments. F.-K.H., X.-Y.X., H.Y., Y.-F.L., and Y.-Q.P. designed, simulated, laid out and packaged the photonic processor. F.-K.H., X.-Y.X., T.-Y.Z., L.F., C.-H.W., J.M., Z.-F.L., and C.-Q.L. conducted the measurements and analyzed the data. F.-K.H., X.-Y.X., and X.-M.J. wrote the paper with input from all the other authors.}

\subsection*{Competing interests}
The authors declare no competing interests.

\clearpage

\clearpage

\clearpage

\clearpage

\clearpage

\clearpage

\end{document}